\documentclass[conference]{IEEEtran}

\usepackage{cite}
\usepackage{amsmath,amssymb,amsfonts}
\usepackage{graphicx}
\usepackage{xcolor}
\usepackage[whole]{bxcjkjatype}
\usepackage{xurl}
\usepackage{booktabs}
\usepackage{colortbl}
\usepackage{eso-pic}

\AddToShipoutPictureBG*{
  \AtPageUpperLeft{
    \raisebox{-1.2cm}{\hspace{2cm}\fbox{\parbox{0.95\textwidth}{\small This paper was originally presented at IEEE CCNC 2025. An extended version has been published in Computers \& Security.\\
    For citation, please refer to the Computers \& Security version: D.~Chiba, H.~Nakano, and T.~Koide, ``DomainDynamics: Advancing Lifecycle-Based Risk Assessment of Domain Names,'' Computers \& Security, 2025, https://doi.org/10.1016/j.cose.2025.104366.}}}
  }
  \AtPageLowerLeft{
    \raisebox{1.2cm}{\hspace{2cm}\parbox{0.95\textwidth}{\footnotesize © 2025 IEEE. Personal use of this material is permitted. Permission from IEEE must be obtained for all other uses, in any current or future media, including reprinting/republishing this material for advertising or promotional purposes, creating new collective works, for resale or redistribution to servers or lists, or reuse of any copyrighted component of this work in other works.}}
  }
}

\begin{document}

\title{DomainDynamics: Lifecycle-Aware Risk Timeline Construction for Domain Names}

\author{
\IEEEauthorblockN{Daiki Chiba, Hiroki Nakano, and Takashi Koide}
\IEEEauthorblockA{NTT Security Holdings Corporation \& NTT Corporation, Tokyo, Japan \\ Email: daiki.chiba@ieee.org}
}

\maketitle

\begin{abstract}
The persistent threat posed by malicious domain names in cyber-attacks underscores the urgent need for effective detection mechanisms. Traditional machine learning methods, while capable of identifying such domains, often suffer from high false positive and false negative rates due to their extensive reliance on historical data. Conventional approaches often overlook the dynamic nature of domain names, the purposes and ownership of which may evolve, potentially rendering risk assessments outdated or irrelevant. To address these shortcomings, we introduce DomainDynamics, a novel system designed to predict domain name risks by considering their lifecycle stages. DomainDynamics constructs a timeline for each domain, evaluating the characteristics of each domain at various points in time to make informed, temporal risk determinations. In an evaluation experiment involving over 85,000 actual malicious domains from malware and phishing incidents, DomainDynamics demonstrated a significant improvement in detection rates, achieving an 82.58\% detection rate with a low false positive rate of 0.41\%. This performance surpasses that of previous studies and commercial services, improving detection capability substantially.
\end{abstract}


\section{Introduction}
\label{sec:introduction}
Malicious domain names utilized in cyber-attacks, such as phishing and malware, remain a persistent threat.
Detecting these fundamentally malicious domains is crucial for cyber defense.
Numerous studies have thus proposed using machine learning (ML) to detect malicious domains.
For example, many practices involve learning the characteristics of malicious domains from a large body of related information to ascertain whether a new domain name is likely to be malicious.

While these approaches are reported to identify domain names involved in malicious activities, they are not without limitations.
Since these ML methods depend on extensive historical data, domains once deemed malicious may still be classified as such, even if their registration has lapsed or they have been rendered harmless, leading to false positives.
Conversely, domains previously employed for legitimate purposes that are later misused may not be promptly identified as malicious, resulting in false negatives.
This is because these methods make determinations based on historical data without considering the specific point in time at which the assessment is made.

To overcome these challenges, this study introduces DomainDynamics, a system capable of assessing the risk of domain names at any given point in their lifecycle.
DomainDynamics constructs a timeline for each domain name, extracts characteristics at various points in time, and employs ML to determine risk based on these temporal features.

Implementing DomainDynamics significantly boosts cyber defense capabilities for Security Operations Centers (SOCs) and Computer Security Incident Response Teams (CSIRTs) by cutting down on false positives and false negatives.
It streamlines the operational efficiency of SOCs and CSIRTs, enabling them to focus on genuine threats rather than wasting resources on false alarms or missing real dangers.
By ensuring a high accuracy in threat detection and minimizing oversight, DomainDynamics empowers these teams to preemptively tackle potential cyber-attacks, enhancing both the security posture and response effectiveness without unnecessary disruption to users.
In essence, the system's adeptness at discerning true cyber threats from benign activity is fundamental in fortifying cyber defense efforts.

In an evaluation experiment involving over 85,000 malicious domains from malware and phishing incidents over 18 months, DomainDynamics accurately predicted domain names that would be used in attacks within 7 days with a detection rate of 82.58\%, while maintaining a low false positive rate of 0.41\%.
This represents a significant improvement---10 times the detection rate of 8.14\% in a previous study.

\begin{table*}[!t]
    \caption{Comparison of detection methods in major studies on malicious domain name detection from 2010 to 2023.}
    \label{table:list-related}
    \centering
    \scriptsize
    \tabcolsep=1.3mm
    {\renewcommand\arraystretch{1.1}
    \begin{tabular}{llll | lll | lllll | ll}
    \toprule
    &  &  &  & Input &  &  & Features &  &  &  &  & Output &  \\
    Literature & Venue & Year & Target & Domain & Path & Content & Lexical & Context & Resource & User & Change & Point & Change \\
    \midrule
    \rowcolor{gray!10}Notos~\cite{DBLP:conf/uss/AntonakakisPDLF10} & Security & 2010 & Malware Domains & \checkmark & - & - & \checkmark & - & \checkmark & - & - & \checkmark & - \\
    EXPOSURE~\cite{DBLP:conf/ndss/BilgeKKB11} & NDSS & 2011 & Malicious Domains & \checkmark & - & - & \checkmark & - & \checkmark & \checkmark & - & \checkmark & - \\
    \rowcolor{gray!10}Szurdi et al.~\cite{DBLP:conf/uss/SzurdiKCSFK14} & Security & 2014 & Typosquatting Domains & \checkmark & - & \checkmark & \checkmark & \checkmark & \checkmark & - & - & \checkmark & - \\
    DomainProfiler~\cite{DBLP:conf/dsn/ChibaYASYMG16} & DSN & 2016 & Malicious Domains & \checkmark & - & -  & \checkmark & - & \checkmark & - & \checkmark & \checkmark & - \\
    \rowcolor{gray!10}Kintis et al.~\cite{DBLP:conf/ccs/KintisMLCGPNA17} & CCS & 2017 & Combosquatting Domains & \checkmark & - & - & \checkmark & - & - & - &-  & \checkmark & - \\
    DomainChroma~\cite{DBLP:journals/compsec/ChibaAYHMG18} & COSE & 2018 & Compromised Domains & \checkmark & - & - & \checkmark & - & \checkmark & - & - & \checkmark & - \\
    \rowcolor{gray!10}FANCI~\cite{DBLP:conf/uss/SchuppenTHM18} & Security & 2018 & Algorithmically Generated Domains & \checkmark & - & - & \checkmark & - & - & - & - & \checkmark & - \\
    DomainScouter~\cite{DBLP:conf/raid/0001HKSGA19} & RAID & 2019 & Homograph \& Combosquatting IDNs & \checkmark & - & - & - & - & \checkmark & - & - & \checkmark & - \\
    \rowcolor{gray!10}COMAR~\cite{DBLP:conf/eurosp/MaroofiKHAD20} & EuroS\&P & 2020 & Compromised Domains & \checkmark & - & \checkmark & \checkmark & \checkmark & \checkmark & \checkmark & - & \checkmark & - \\
    De Silva et al.~\cite{DBLP:conf/uss/SilvaNEKYK21} & Security & 2021 & Compromised Domains & \checkmark & \checkmark & - & \checkmark & - & \checkmark & - & - & \checkmark & - \\
    \rowcolor{gray!10}Sabah et al.~\cite{DBLP:conf/raid/SabahNBC22} & RAID & 2022 & Phishing Domains & \checkmark & - & - & \checkmark & - & \checkmark & - & \checkmark & \checkmark & - \\
    PhishReplicant~\cite{DBLP:conf/acsac/KoideFN023} & ACSAC & 2023 & Generated Squatting Domains & \checkmark & - & - & \checkmark & - & - & - & - & \checkmark & -\\
    \midrule
    \rowcolor{yellow!25}\textbf{DomainDynamics} & - & - & \textbf{Malware \& Phishing Domains} & \checkmark & - & - & - & - & \checkmark & - & \checkmark & \checkmark & \checkmark \\
    \bottomrule
    \end{tabular}
    }
\end{table*}

\section{Background and Related Work}
\label{sec:background}
This section organizes past research on malicious domain names and explains the challenges this study aims to address, using motivating examples.

\subsection{Detecting Malicious Domains}
\label{sec:detecting-malicious}
Table~\ref{table:list-related} summarizes 12 key studies on malicious domain detection (2010---2023), detailing venue, year, targets, inputs, ML features, and outputs.
Based on the table, we will now describe each study in order based on their target, specifically what they aim to detect within malicious domain names.

\noindent\textbf{Malware Domains.}
Malware domains, essential for command-and-control (C2) operations and malware dissemination, represent the field's earliest research focus, with studies leveraging ISP-level DNS data to evaluate them, including the dynamic DNS reputation system like Notos~\cite{DBLP:conf/uss/AntonakakisPDLF10}.

\noindent\textbf{Algorithmically Generated Domains.}
Algorithmically generated domains (AGDs) are domain names generated by domain generation algorithms (DGA) within malware domains.
For example, FANCI is a system that extracts features from NXDomain domain names and classifies AGDs and other NXDomains using a ML-based classifier~\cite{DBLP:conf/uss/SchuppenTHM18}.

\noindent\textbf{Phishing Domains.}
Research on phishing domains has evolved to address the growing threat of online fraud. Recent studies have employed diverse data sources and techniques for detection, including Certificate Transparency (CT) logs~\cite{DBLP:conf/raid/SabahNBC22}.

\noindent\textbf{Squatting Domains.}
Squatting domains mimic legitimate ones, with studies focusing on detection and characteristics.
Squatting domains sometimes include phishing domains, but here we focus on those specifically related to squatting.
First, there are studies on typosquatting, which can be caused by typographical errors on the keyboard~\cite{DBLP:conf/uss/SzurdiKCSFK14}.
Next, there are multiple studies reporting on combosquatting and similar situations where deception is based on visual appearance by adding keywords to legitimate brand names~\cite{DBLP:conf/ccs/KintisMLCGPNA17}.
Moreover, recent studies have reported on detection methods and survey results for homograph internationalized domain names (IDN)~\cite{DBLP:conf/raid/0001HKSGA19}.
There is also a very recent study that first reported on generated squatting domains, which are considered a combination of multiple squatting methods~\cite{DBLP:conf/acsac/KoideFN023}.

\noindent\textbf{Compromised Domains.}
Compromised domains, initially legitimate, are later misused by attackers.
DomainChroma is the first paper to mention the need to distinguish whether a domain name is compromised as a separate axis from being malicious, producing practical blocklists from malicious domains~\cite{DBLP:journals/compsec/ChibaAYHMG18}.
The same concept has since been proposed in other papers, each as a system called COMAR~\cite{DBLP:conf/eurosp/MaroofiKHAD20}, and as a separate ML system~\cite{DBLP:conf/uss/SilvaNEKYK21}.

\noindent\textbf{Malicious Domains.}
Malicious domain research broadly targets various domains, not limited to specific categories.
EXPOSURE~\cite{DBLP:conf/ndss/BilgeKKB11} and DomainProfiler~\cite{DBLP:conf/dsn/ChibaYASYMG16} are representative studies that perform so-called domain reputation by detecting general malicious domains using DNS data and ML.

\subsection{Motivating Examples}
\label{sec:motivating}
We present two domain names as motivating examples to illustrate the challenges this study seeks to overcome.
The first is a domain name previously involved in malicious activities but has since changed ownership, ceased to be involved in attacks, and had its DNS records deleted, rendering it currently inactive.
The second is a domain name that once offered legitimate services but was later captured through drop catching, changed ownership, and is now being used for malicious purposes.
The studies listed in Table~\ref{table:list-related} lack open access for real-time risk assessment of current domain names.
Therefore, we use VirusTotal to obtain risk assessments for these examples.
VirusTotal provides the latest risk assessments for domain names from 89 different security engines and is widely referenced in both industry and academia.
The first domain name, now inaccessible worldwide, should pose no risk, yet VirusTotal results, particularly from AI-powered engines, labeled it as Malicious.
This indicates that past evaluations have a strong, lingering influence.
The second domain name, actively used for malicious purposes, should be considered risky, but VirusTotal's engines mostly assessed it as Clean.
These examples lead us to the following research question:

\noindent\textit{RQ: How can considering the lifecycle changes of domain names enhance the accuracy of their risk assessments?}

As demonstrated by these examples, domain names can differ significantly over time in terms of registrants, DNS settings, and web content.
Thus, a single risk assessment at one point---especially the most recent---may not suffice for an accurate evaluation.
Unlike image classification tasks, where a dog remains classified as a dog, or binary malware classification, where malware retains its label, domain names require risk assessments that adapt over time.
To draw an analogy, it would be akin to evaluating a completely different restaurant with a new owner based on reviews of a restaurant that has closed.
The core issue is that judgments are made without considering the timing of the results---whether they pertain to the era of the previous establishment or that of the new one.

\section{Proposed System: DomainDynamics}
\label{sec:proposed-system}

\subsection{Goal and Scope}
\label{sec:goal-scope}
This research primarily aims to develop a system capable of assessing the risk of malicious domain names by considering their temporal context, thus enabling risk determinations at any given moment, from the past to the future.
As outlined in Section~\ref{sec:motivating}, this system is designed to reduce incidents by preventing continuous false positive detections for domain names that were once malicious but are no longer active.
In contrast, it can swiftly identify a domain name that has transitioned from being used for legitimate services to being exploited for malicious purposes, thereby preventing false negatives.

For an overview of the scope of this research and its relationship with previous studies, refer to Table~\ref{table:list-related}.

\noindent\textbf{Target.}
Previous research has generally focused on malicious domains, with some studies specifically targeting malware domains, algorithmically generated domains, phishing domains, typosquatting domains, combosquatting domains, and homograph IDNs, among others.
In contrast, this research specifically focuses on malware and phishing domains.

\noindent\textbf{Input.}
Regarding the input to the system, most past studies have concentrated on domain names, though some have also utilized the full path of URLs or content such as text or screenshots from websites.
In this research, the input is limited solely to domain names.

\noindent\textbf{Features.}
A key distinction of this research from previous work is the combination of features used in ML.
Table~\ref{table:list-related} categorizes features into five groups: Lexical, Context, Resource, User, and Change.
Lexical features originate from the character strings of domain names themselves, including n-grams, domain name length, and the entropy of the domain name string.
Context features derive from content accessed via the web, such as website text, screenshots, or logo images.
Resource features pertain to information related to domain names, such as WHOIS records, IP addresses, DNS records, TLS certificates, and web server data.
User features are derived from user interactions and communications, such as the domain names accessed through an organization's proxy server or data from social media posts.
Change features relate to long-term changes in domain-related information, such as the transition of inclusion in Top Lists (DomainProfiler~\cite{DBLP:conf/dsn/ChibaYASYMG16}).
This research employs features from both Resource and Change categories to enable risk assessment throughout the entire lifecycle of a domain name.
This approach is chosen because reliance on content, such as the web-accessible content of domain names, can hinder risk prediction if the content is cloaked or if it is not feasible to monitor and follow all content changes for each domain name in a scalable manner.

\noindent\textbf{Output.}
The output of most prior research is limited to a binary classification indicating whether a domain name is malicious at a specific point in time.
In contrast, our research presents a significant advancement: the generation of a continuous risk timeline.
This timeline not only provides the risk assessment at the evaluation moment but also predicts the risk evolution from the past into the upcoming future for each domain name.
Unlike the static point-in-time analysis commonly found in existing literature, our system dynamically models the risk associated with domain names, reflecting changes in their lifecycle.
This continuous risk timeline offers a more nuanced understanding of domain risk, accommodating the fluid nature of domain usage and threat levels over time.

\noindent\textbf{Summary.}
Refer to Table~\ref{table:list-related} for a comparison of DomainDynamics with prior research.
While previous studies have incorporated temporal changes as features (refer to the Change column in Table~\ref{table:list-related}), none have demonstrated the temporal risk variations for individual domain names as achieved by DomainDynamics (refer to the Output column in Table~\ref{table:list-related}).
Additionally, DomainDynamics is unique in its focus on malware and phishing domains (see the Target column in Table~\ref{table:list-related}) and is distinguished by its use of a combination of resource and change features, independent of content (see the Features column in Table~\ref{table:list-related}).
\textit{In summary, DomainDynamics stands apart from conventional research in terms of both the features it uses and the outputs it generates.}

\begin{figure}[!t]
    \centering
    \includegraphics[width=\linewidth]{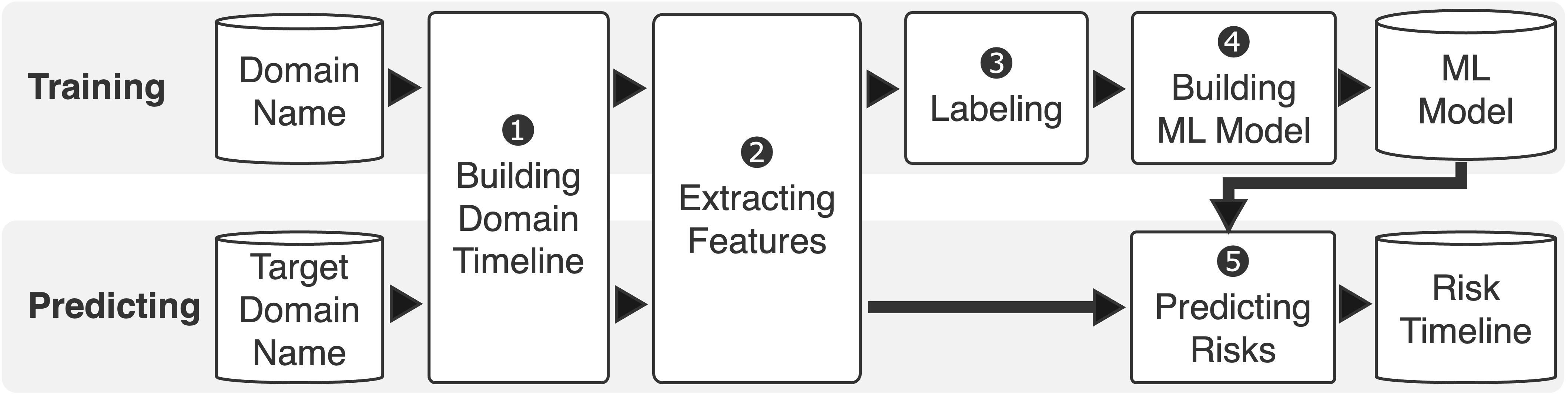}
    \caption{DomainDynamics system overview.}
    \label{fig:system}
\end{figure}

\subsection{System Overview}
DomainDynamics, the system proposed in this paper, is designed to achieve the goals and scope outlined above.
Figure~\ref{fig:system} presents a diagrammatic overview of the system.
DomainDynamics operates in two distinct phases: Training and Predicting.

During the Training phase, domain names are input.
The process begins with the construction of a Domain Timeline (❶), followed by the extraction of Features for each point on the Domain Timeline (❷).
Using ground truth data, the points in time when the domain was involved in attacks are determined, and Labels are assigned accordingly (❸).
Subsequently, an ML Model is trained (❹).

In the Predicting phase, a Target Domain Name is input, and the Domain Timeline and Features are processed similarly to the Training phase (❶, ❷).
The ML Model trained earlier is then used to predict the points at which the Target Domain Name might be exploited for attacks (❺).

The subsequent sections detail each step of the process.

\subsection{❶ Building Domain Timeline}
This step is fundamental to both the Training and Predicting phases.
A domain name is input, and a historical database pertinent to the domain name is referenced to construct a Domain Timeline.
The Domain Timeline represents the lifecycle changes of a domain name, capturing its evolution over time.

To track these lifecycle changes, the study utilizes historical databases of WHOIS records, SOA records, and TLS certificates.
For WHOIS records, changes in Registrar, Creation Date, Expiry Date, and Domain Status are monitored.
In SOA records, alterations in MNAME, RNAME, and SERIAL are tracked.
For TLS certificates, especially pertinent for Web/HTTPS use, changes in Issuer C, Issuer CN, Issuer O, Validity Not Before, Validity Not After, and Subject CN are followed.

It is important to note that the construction of a comprehensive historical database is beyond the scope of this study; instead, commercial or paid databases were used for WHOIS records, SOA records, and TLS certificates data.
These databases should be frequently updated, ideally on a daily basis, to ensure that the most recent and accurate information is always available for constructing the Domain Timeline.

\begin{figure}[!t]
    \centering
    \includegraphics[width=\linewidth]{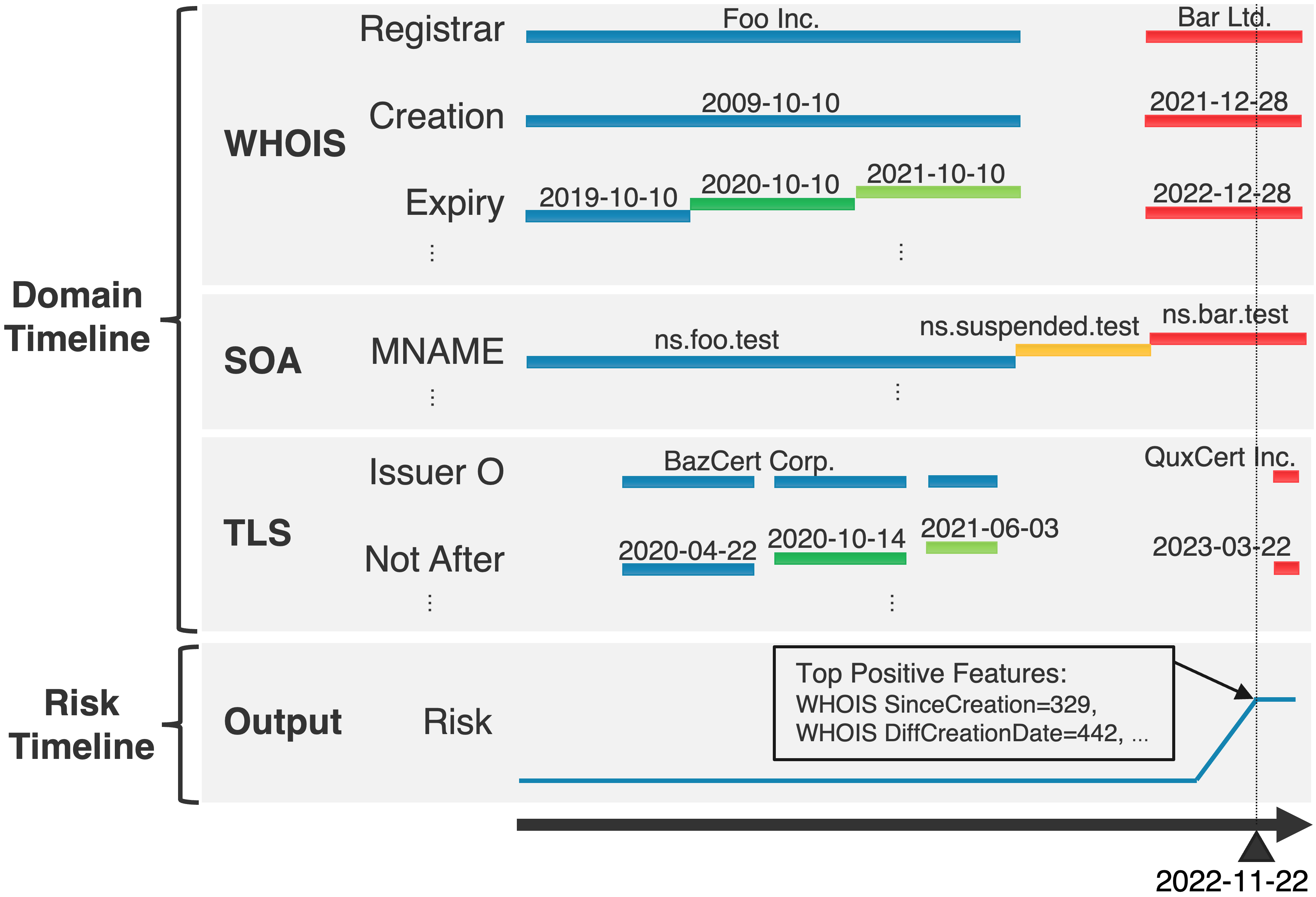}
    \caption{Domain Timeline (top) and Risk Timeline (bottom).}
    \label{fig:timeline}
\end{figure}

A schematic representation of the Domain Timeline for a domain name is depicted in the upper part of Figure~\ref{fig:timeline}.

The domain name featured in Figure~\ref{fig:timeline} had been used for legitimate purposes for over 10 years but was captured by an attacker, left idle for about a year to avoid easy detection, and then abruptly put to malicious use, based on an actual case.
In the WHOIS Domain Timeline in Figure~\ref{fig:timeline}, we can see that the Registrar changed from \texttt{Foo Inc} to \texttt{Bar Ltd} at a certain point, the Creation Date, which had been maintained at \texttt{2009-10-10}, expired and was re-registered on \texttt{2021-12-28}, and the Expiry Date, which had been extended annually from \texttt{2019-10-10} to \texttt{2021-10-10}, then expired.
Additionally, the SOA Domain Timeline in Figure~\ref{fig:timeline} shows that MNAME changed from \texttt{ns.foo.test} to \texttt{ns.suspended.test}, and later to \texttt{ns.bar.test}.
Furthermore, the TLS Domain Timeline in Figure~\ref{fig:timeline} indicates transitions in Issuer O and Not After, revealing that during the period when \texttt{Foo Inc} was the Registrar, the domain started using certificates from \texttt{BazCert Corp}, but there were periods of certificate expiration leading to gaps, and the certificates expired before the domain itself did, and after the Registrar changed to \texttt{Bar Ltd}, it took some time before certificates from \texttt{QuxCert Inc} began to be activated.

\subsection{❷ Extracting Features}
\label{sec:extracting-features}
This step is integral to both the Training and Predicting phases, where features of domain names are extracted at each point in time by referring to the Domain Timeline.
In essence, features for a single domain name are extracted at multiple temporal points.
These features will later serve as attributes to predict the future risk associated with a domain name at a specific point in time.
Table~\ref{tab:features} enumerates the features extracted from the Domain Timeline.
The intuition behind using WHOIS, SOA, and TLS records as feature sources lies in their ability to encapsulate different aspects of a domain's lifecycle.
WHOIS Records provide a foundational glimpse into the domain's registration and administrative lineage, crucial for identifying ownership changes or inactivity that may signal misuse.
SOA Records reflect operational readiness within the DNS hierarchy, marking key lifecycle events that could indicate a domain's transition to active use or susceptibility to malicious exploitation.
TLS Certificates signal a domain's readiness for secure online operations, with changes in certificates often preceding the active deployment of web services, including those for nefarious purposes.
At a designated date in the Domain Timeline, features are generated solely from data available prior to that point in time, from the perspective of WHOIS records, SOA records, and TLS certificates.

\begin{table*}[!t]
    \caption{Features extracted from domain historical data.}
    \label{tab:features}
    \centering
    \begin{tabular}{l r l r}
    \toprule
    Category & No. & Feature & Dimension\\
    \midrule
    ① WHOIS Records & 1 & Unique WHOIS records count & 4 \\
    & 2 & Days between WHOIS dates & 3 \\
    & 3 & Domain status & 14 \\
    & 4 & Days between WHOIS updates & 4 \\
    & 5 & Days difference between WHOIS records & 3 \\
    & 6 & WHOIS records change & 2 \\
    \midrule
    ② SOA Records & 7 & Unique SOA records count & 3 \\
    & 8 & Days between SOA dates & 1 \\
    & 9 & Days between SOA updates & 4 \\
    & 10 & Days difference between SOA records & 1 \\
    & 11 & SOA records change & 2\\
    \midrule
    ③ TLS Certificates & 12 & Unique TLS certificates count & 6 \\
    & 13 & Days between TLS dates & 2 \\
    & 14 & Days between TLS updates & 4 \\
    & 15 & Days difference between TLS certificates & 2\\
    & 16 & TLS certificates change & 4\\
    \bottomrule
    \end{tabular}
\end{table*}

\subsubsection{① WHOIS Records}\mbox{}\\
\noindent\textbf{No. 1: Unique WHOIS records count.}
Variations in the registrar or domain name status can indicate changes in the domain's lifecycle.
Features are thus designed to effectively capture such histories.
Specifically, the count of unique values historically registered for items such as Registrar, Creation Date, Expiry Date, and Domain Status in the WHOIS records is employed as a feature.
For example, at the point in time of 2022-11-22 in Figure~\ref{fig:timeline}, there are two unique values registered historically for Registrar, Foo Inc and Bar Ltd, resulting in a Unique WHOIS records count of 2.

\noindent\textbf{No. 2: Days between WHOIS dates.}
These features are designed to quantify the time elapsed since lifecycle changes at each feature extraction point.
Specifically, the number of days from dates such as the Creation Date, Expiry Date, and Updated Date in the WHOIS records to the point in time is used as a feature.

\noindent\textbf{No. 3: Domain status.}
At each feature extraction point, the most recent domain status (e.g., clientTransferProhibited) from the WHOIS record is used as a feature to determine whether the domain name is in a state that is normatively usable from a lifecycle perspective.

\noindent\textbf{No. 4: Days between WHOIS updates.}
These features are designed to leverage lifecycle changes by assessing the frequency of WHOIS record updates up to each feature extraction point.
Specifically, the minimum, mean, median, and maximum number of days between updates across multiple historical WHOIS records are used as features.

\noindent\textbf{No. 5: Days difference between WHOIS records.}
To discern whether WHOIS changes are routine updates for legitimate purposes, the difference in days between the dates (Updated Date, Creation Date, Expiry Date) of the most recent WHOIS record and the preceding one is used as a feature at each feature extraction point.

\noindent\textbf{No. 6: WHOIS records change.}
To ascertain whether the most recent WHOIS change was for legitimate reasons, a boolean feature is employed at each feature extraction point to determine whether the Registrar and Domain Status of the most recent WHOIS record differ from the previous one.

\subsubsection{② SOA Records}\mbox{}\\
\noindent\textbf{No. 7: Unique SOA records count.}
Alterations in the DNS SOA records can signify lifecycle shifts for a domain name.
Features are designed to effectively capture such histories.
Specifically, the number of distinct values historically registered for items such as MNAME, RNAME, and SERIAL in the SOA records is utilized as a feature.

\noindent\textbf{No. 8: Days between SOA dates.}
This feature helps understand the time elapsed since lifecycle changes at each feature extraction point.
Specifically, the number of days from the last updated date of SOA record to the point in time is used as a feature.

\noindent\textbf{No. 9: Days between SOA updates.}
These features leverage lifecycle changes by determining the frequency of SOA record updates up to each feature extraction point.
Specifically, the minimum, mean, median, and maximum number of days between updates across multiple historical SOA records are used as features.

\noindent\textbf{No. 10: Days difference between SOA records.}
To distinguish whether SOA changes are regular updates for legitimate purposes, the difference in days between the dates of the most recent SOA record and the previous one is used as a feature at each feature extraction point.

\noindent\textbf{No. 11: SOA records change.}
To identify whether the most recent SOA change was for legitimate reasons, a boolean feature is used at each feature extraction point to determine whether the MNAME and RNAME of the most recent SOA record differ from the preceding one.

\subsubsection{③ TLS Certificates}\mbox{}\\
\noindent\textbf{No. 12: Unique TLS certificates count.}
Shifts indicated by changes in TLS certificates are critical to understanding the domain's lifecycle.
Features are therefore designed to effectively capture these histories.
Specifically, the count of unique values historically registered for items such as Issuer C, Issuer CN, Issuer O, Validity Not Before, Validity Not After, and Subject CN in the TLS certificates is used as a feature.

\noindent\textbf{No. 13: Days between TLS dates.}
These features gauge the time elapsed since lifecycle changes at each feature extraction point.
Specifically, the number of days from each date such as Validity Not Before and Validity Not After in the TLS certificates to the point in time is used as a feature.

\noindent\textbf{No. 14: Days between TLS updates.}
These features utilize lifecycle changes by assessing the frequency of TLS record updates up to each feature extraction point.
Specifically, the minimum, mean, median, and maximum number of days between updates across multiple historical TLS certificates are used as features.

\noindent\textbf{No. 15: Days difference between TLS certificates.}
To determine whether TLS changes are routine updates for legitimate purposes, the difference in days between the dates (Validity Not Before, Validity Not After) of the most recent TLS record and the one prior is used as a feature at each feature extraction point.

\noindent\textbf{No. 16: TLS certificates change.}
To discern whether the most recent TLS change was for legitimate reasons, a boolean feature is employed at each feature extraction point to determine whether the Issuer C, Issuer CN, Issuer O, and Subject CN of the most recent TLS record differ from the previous one.

\subsection{❸ Labeling}
\label{sec:labeling}
Labeling is a step exclusive to the training phase, wherein ground truth labels are assigned to indicate the risk associated with domain names at each point in time.
The original purpose of DomainDynamics was to predict the likelihood of a domain name being used for an attack in advance.

In this study, ground truth labels denote whether a domain name will be used for an attack within $N$ days following a specific point in time.
For each temporal point extracted in the Feature Extraction step, if there is verified evidence that the domain name was used for an attack within the subsequent $N$ days, a Malicious/Positive label is assigned; otherwise, a Non-malicious/Negative label is applied.
The parameter $N$ is adjustable and will be evaluated in detail in Section~\ref{sec:detection-performance-parameters}.

For instance, if $N$ is set to 30 and it is known that a domain name hosted a phishing site on January 30, 2024, then features of that domain name prior to January 1, 2024, would receive a Non-malicious/Negative label, as there was no attack within the following 30 days.
Conversely, features after January 1, 2024, would be labeled Malicious/Positive because the domain would be used for an attack within the next 30 days.

\subsection{❹ Training ML Model}
This step, which is part of the training process, involves creating a supervised ML model using the features and labels of domain names associated with each time point.
In supervised ML, a model is initially trained with known data (in this case, the features and labels at each time point), and then this model is used to predict the risk for new, unseen data (in this case, the features of the Target Domain Name).

Several algorithms are suitable for supervised ML; given that the problem is defined with temporal features for domain names and binary labels (Positive/Negative), any binary classification algorithm is applicable.
The performance of various algorithms will be assessed in Section~\ref{sec:detection-performance-parameters}.

\subsection{❺ Predicting Risks}
\label{sec:predicting-risks}
During the prediction phase, the ML Model trained earlier is used to estimate the risk of the Target Domain Name being exploited in an attack within the next $N$ days.
This estimation is based on the features at each time point for the Target Domain Name.
The risk at each point is represented by a prediction probability generated by the ML Model.

The prediction probability is a continuous value ranging from 0 to 1, with values closer to 0 suggesting Non-malicious/Negative and those closer to 1 indicating Malicious/Positive.
The risk for each time point of the Target Domain Name is presented as a Risk Timeline, as depicted at the bottom of Figure~\ref{fig:timeline}.

In addition to predicting the risk probability, this step incorporates techniques from XAI (eXplainable Artificial Intelligence) to elucidate the features influencing each probability prediction.
In this research, SHAP (SHapley Additive exPlanations)~\cite{DBLP:conf/nips/LundbergL17} is utilized as the XAI method.
SHAP quantifies the contribution of each input feature to the prediction probability provided by the ML Model, assigning SHAP Values that reflect the impact of each feature.

Using SHAP Values, the top contributing features, both positive and negative, are identified for each risk prediction at every time point.
For example, as demonstrated in the Risk Timeline of Figure~\ref{fig:timeline}, hovering over the prediction probability at a specific time point reveals the top contributing features that influenced the probability at that moment.

\section{Evaluation}
\label{sec:evaluation}
In this section, we conduct a performance evaluation to answer the following three questions, assessing the effectiveness of DomainDynamics:

\noindent\textit{Q1: Can DomainDynamics predict the risk of a domain name being used in an attack before it occurs?}

\noindent\textit{Q2: What are the intrinsic characteristics of DomainDynamics?}

\noindent\textit{Q3: How does DomainDynamics compare to baseline systems in terms of detection capability?}

To address Q1, we compiled a dataset of domain names that were actually used in attacks and evaluated our system's predictive performance.
For Q2, we explore the impact of system-required parameters and the influence of features on predictions, characterizing DomainDynamics.
In response to Q3, we introduce new evaluation metrics that account for the temporal change in the risk of domain names and present a comparative performance analysis of DomainDynamics against two baseline systems.

\subsection{Metrics}
\label{sec:metrics}
We introduce new evaluation metrics that consider the temporal changes in domain name risk, enabling a more nuanced understanding of detection capabilities.
We redefine and utilize the following four metrics:

\noindent\textbf{True Positive (TP):}
A case is considered a True Positive if a malicious domain name, actually used in an attack (e.g., malware, phishing), is detected before the attack takes place.

\noindent\textbf{False Negative (FN):}
A case is considered a False Negative if a malicious domain name, actually used in an attack, is not detected before the attack takes place.

\noindent\textbf{False Positive (FP):}
A case is considered a False Positive if a benign domain name, not used in an attack, is incorrectly identified as malicious at least once during the observation period.

\noindent\textbf{True Negative (TN):}
A case is considered a True Negative if a benign domain name, not used in an attack, is never falsely identified as malicious during the observation period.

Additionally, we define the following four ratio-based metrics: False Positive Rate (FPR), Precision, Recall (also known as True Positive Rate, TPR), and F1 Score.

The rationale for establishing such metrics is to accurately assess detection performance considering the temporal dynamics of domain name risk.
In traditional studies, as documented in Table~\ref{table:list-related}, domain names collected over a certain period using blocklists and similar methods are labeled ``Malicious,'' even if they are no longer active in attacks thereafter.
Counting predictions of these domain names as TPs can lead to an inflated detection rate.
Furthermore, our new metrics are particularly stringent for DomainDynamics, as any single error made during the observation period results in an FP or FN.

\begin{table*}[!t]
    \caption{Ground truth datasets.}
    \label{tab:dataset}
    \centering
    \begin{tabular}{llllr}
    \toprule
    Dataset & Ground Truth  & For Training & For Predicting & \# FQDNs \\
    \midrule
    Malware & Malicious & \checkmark & \checkmark & 3,410 \\
    Phishing & Malicious & \checkmark & \checkmark & 30,711 \\
    CrowdCanary & Malicious & \checkmark & \checkmark & 51,502 \\
    Tranco Top10k & Non-malicious &  & \checkmark & 4,358 \\
    \bottomrule
    \end{tabular}
\end{table*}

\subsection{Dataset}
To accurately evaluate the system using the new metrics described previously, a ground truth dataset is essential.
Such a dataset must accurately reflect periods when a given domain name was not used for attacks and when it began to be used for attacks.
Therefore, in this study, we carefully prepared three types of malicious ground truth datasets and one type of benign ground truth dataset.
The details of each dataset are summarized in Table~\ref{tab:dataset}.

\noindent\textbf{Malware.}
The malware dataset comprises domain names that were confirmed to function as command and control (C2) for malware at the time of observation, indicating when a C2 domain name was active.
This dataset was obtained through a partnership with a managed security service provider and includes unique C2 fully qualified domain names (FQDNs) of 47 malware families, collected over 18 months from April 2022 to September 2023, totaling 3,410 entries.
While we cannot disclose all details of this dataset, it is based on information used to protect customers in a commercial setting.

\noindent\textbf{Phishing.}
The phishing dataset contains domain names that were active as phishing sites at the time of observation.
Specifically, we utilized commercial OpenPhish data and targeted those with evidence of operation, such as screenshots taken at specific timestamps.
This dataset includes 30,711 unique FQDNs over the same 18-month period.

\noindent\textbf{CrowdCanary.}
CrowdCanary is a dataset comprising domain names that users have definitively reported as being used in phishing emails or phishing websites at the time of observation.
CrowdCanary, a system recently proposed in literature~\cite{DBLP:conf/IEEEares/Nakano0KFYHYM23}, monitors text and images posted on platforms like X/Twitter to extract phishing reports, and we received a research dataset from the authors.
We used 42,756 posts collected over 18 months, with the criteria that the tweets were from security experts, included images (indicating that the sites were accessible at the time), and contained URL or domain name information in the tweet text (ensuring high reliability of the information).
Some posts contain multiple domain names, resulting in a total of 51,502 unique FQDNs.

These three malicious datasets will be used for training and evaluating the ML model's predictions.
To prevent temporal leakage---where future data (the ground truth) would be used to predict past data---in the evaluation, we will use data up to March 2023 for training and data from April 2023 to September 2023 for prediction.
It is important to note that there are no overlapping FQDNs between these three datasets, and naturally, no FQDNs overlap between the training and prediction sets.

\noindent\textbf{Tranco Top10k (Non-malicious).}
The ground truth for benign FQDNs consists of domain names that were consistently ranked in the top 10,000 of the Tranco list~\cite{DBLP:conf/ndss/PochatGTKJ19} during the 18 months from April 2022 to September 2023.
To ensure the absence of malicious domains, each domain's website content was manually checked.
These were used only as a reference for evaluation during prediction as FQDNs that were never used for attacks during the period and were not used for training.
As outlined in Section~\ref{sec:labeling}, DomainDynamics considers the lifecycle of a domain name, labeling it as Non-malicious before it is used for an attack and as Malicious when it is used for an attack.
Therefore, it should be emphasized that FQDNs that have always been benign are not necessarily required for training the ML model, and in this case, they are not used.
The Tranco Top10k is prepared solely to evaluate whether DomainDynamics is causing obvious false positives.

\begin{table*}[!t]
    \caption{DomainDynamics with different parameters.}
    \label{tab:performance}
    \centering
    {\renewcommand\arraystretch{1.1}
    \begin{tabular}{llrrrr}
    \toprule
    & Parameter & FPR (↓better) & Precision (↑better) & Recall (↑better)  & F1 (↑better) \\
    \midrule
    Feature Set & WHOIS & 0.96\% & 99.29\% & 35.52\% & 52.32\% \\
    & SOA & 0.60\% & 99.66\% & 46.57\% & 63.48\% \\
    & TLS & 1.88\% & 99.33\% & 73.85\% & 84.71\% \\
    & WHOIS+SOA & \textbf{0.32\%} & 99.85\% & 58.35\% & 73.66\% \\
    & WHOIS+TLS & 0.69\% & 99.78\% & 81.80\% & 89.90\% \\
    & SOA+TLS & 0.78\% & 99.73\% & 77.25\% & 87.06\% \\
    & \cellcolor{yellow!25}\textbf{ALL} & \cellcolor{yellow!25}0.41\% & \cellcolor{yellow!25}\textbf{99.87}\% & \cellcolor{yellow!25}\textbf{82.58\%} & \cellcolor{yellow!25}\textbf{90.40\%} \\
    \midrule
    Labeling Period & \cellcolor{yellow!25}\textbf{7} & \cellcolor{yellow!25}\textbf{0.41\%} & \cellcolor{yellow!25}\textbf{99.87\%} & \cellcolor{yellow!25}82.58\% & \cellcolor{yellow!25}90.40\% \\
    & 14 & 0.73\% & 99.77\% & 82.84\% & 90.52\% \\
    & 30 & 1.10\% & 99.66\% & 85.82\% & 92.23\% \\
    & 60 & 3.49\% & 99.07\% & 97.92\% & 98.49\% \\
    & 90 & 3.53\% & 99.06\% & \textbf{98.24\%} & \textbf{98.64\%} \\
    \midrule
    Machine Learning Algorithm & Decision Tree & 9.59\% & 96.93\% & 80.35\% & 87.86\% \\
    & Random Forest & 1.93\% & 99.38\% & 81.89\% & 89.79\% \\
    & LightGBM & 1.24\% & 99.60\% & 82.21\% & 90.08\% \\
    & \cellcolor{yellow!25}\textbf{XGBoost} & \cellcolor{yellow!25}\textbf{0.41\%} & \cellcolor{yellow!25}\textbf{99.87\%} & \cellcolor{yellow!25}\textbf{82.58\%} & \cellcolor{yellow!25}\textbf{90.40\%} \\
    \bottomrule
    \end{tabular}
    }
\end{table*}

\subsection{Performance with Variable Parameters}
\label{sec:detection-performance-parameters}
We evaluated the performance of DomainDynamics by varying parameters adjustable within the system, utilizing the metrics mentioned in Section~\ref{sec:metrics}.
The parameters in question include the Feature Set, Labeling Period, and ML Algorithm.
Due to space constraints, we report only the results of evaluations with other parameters held constant, while varying the target parameter.

\noindent\textbf{Feature Set.}
As explained in Section~\ref{sec:extracting-features}, DomainDynamics prepares three types of feature sets: WHOIS records, SOA records, and TLS certificates.
We evaluated the difference in performance when using combinations of these feature sets.
Table~\ref{tab:performance} summarizes the detection performance when changing the feature set to WHOIS, SOA, TLS, WHOIS+SOA, WHOIS+TLS, SOA+TLS, and WHOIS+SOA+TLS (ALL).
For evaluation metrics, a lower FPR is preferable, while higher values in Precision, Recall/TPR, and F1 are desirable.
It should be noted that Precision and Recall/TPR are generally in a trade-off relationship, so it may not be possible to optimize both simultaneously.
From the perspective of FPR and Precision, the combination of WHOIS+SOA shows the best results with 0.32\% and 99.85\%, respectively.
However, when using WHOIS+SOA, the Recall/TPR is relatively low at 58.35\%, resulting in an F1 of only 73.66\%, indicating a modest proportion of malicious domain names detected before an attack begins.
When using WHOIS+SOA+TLS (ALL), the Recall/TPR is the highest at 82.58\%, resulting in the highest F1 of 90.40\%.
Although the FPR is higher compared to WHOIS+SOA, it remains a low value of 0.41\%, leading us to conclude that WHOIS+SOA+TLS (ALL) is the most effective parameter set.

\noindent\textbf{Labeling Period.}
We evaluated the performance of DomainDynamics by varying the labeling period $N$ as per the procedure in Section~\ref{sec:labeling}.
This serves to investigate how well the system performs in predicting future risks.
Table~\ref{tab:performance} summarizes the detection performance when changing $N$ to 7, 14, 30, 60, and 90 days.
The results show that at $N=7$, the FPR is the lowest at 0.41\% and the Precision is the highest at 99.87\%.
Increasing $N$ results in a higher FPR, which is expected as a longer forecasting period increases the likelihood of making incorrect predictions.
Conversely, for Recall/TPR and F1, except for the case of $N=14$, the values increase as $N$ increases because a longer forecasting period provides more opportunities to detect malicious domain names before attacks begin.
This result indicates a trade-off, and the choice of parameter will be determined by the acceptable FPR.
However, considering the volume of data handled in this study (tens of thousands of cases or more), an FPR of 1\% or more would result in an excessive number of false positives.
Therefore, we select $N=7$ with an FPR of 0.41\% as the optimal parameter.

\noindent\textbf{ML Algorithm.}
We evaluated the performance of DomainDynamics by varying the ML algorithms used for binary classification.
For this study, we selected and compared four types of ML algorithms: Decision Tree, Random Forest, LightGBM, and XGBoost.
We chose these algorithms because they are all tree-based, which means they do not require feature scaling, and we expect them to be robust to differences in feature sets and variations in training data type and period.
Additionally, by selecting algorithms with significantly different characteristics, we can compare their respective features.
Specifically, Decision Tree is a basic tree model, Random Forest is an ensemble learning algorithm combining many decision trees, and LightGBM and XGBoost are sequential tree-building algorithms known as gradient boosting trees.
For adjustable parameters of each selected algorithm, we used Optuna to search for parameters that minimize FPR and selected the best ones.
Table~\ref{tab:performance} summarizes the performance for each ML algorithm.
The results are clear: XGBoost performed best across all metrics, with an FPR of 0.41\%, Precision of 99.87\%, Recall/TPR of 82.58\%, and F1 of 90.40\%, leading us to conclude that XGBoost is the most suitable algorithm for our experimental setup.

\noindent\textbf{Summary.}
In summary, DomainDynamics demonstrates its best performance with the Feature Set WHOIS+SOA+TLS (ALL), a Labeling Period of 7 days, and the XGBoost ML Algorithm.
In subsequent evaluations, we will employ DomainDynamics trained with these parameters.
It should be noted that we operated DomainDynamics on a standard Linux machine (4-core CPU, 16GB memory) and confirmed that it functions sufficiently fast without the need for a GPU, as it utilizes conventional binary classification ML.
It took on average only 0.73 seconds to input a domain name and output the risk prediction result with the best performance parameter combination and even when running in single-threaded mode without parallelization, proving that it is practically efficient.

\begin{table*}[!t]
    \caption{Performance of DomainDynamics vs Baseline.}
    \label{tab:performance-baselines}
    \centering
    {\renewcommand\arraystretch{1.1}
    \begin{tabular}{llrrrr}
    \toprule
    Predicting Dataset & System & FPR (↓better) & Precision (↑better) & Recall (↑better)  & F1 (↑better) \\
    \midrule
    Malware & \cellcolor{yellow!25}\textbf{DomainDynamics} & \cellcolor{yellow!25}0.41\% & \cellcolor{yellow!25}98.77\% & \cellcolor{yellow!25}\textbf{72.29\%} & \cellcolor{yellow!25}\textbf{83.48\%} \\
    & Baseline & \textbf{0.00\%} & \textbf{100.00\%} & 4.71\% & 9.00\% \\
    \midrule
    Phishing & \cellcolor{yellow!25}\textbf{DomainDynamics} & \cellcolor{yellow!25}0.41\% & \cellcolor{yellow!25}99.70\% & \cellcolor{yellow!25}\textbf{75.73\%} & \cellcolor{yellow!25}\textbf{86.08\%} \\
    & Baseline & \textbf{0.00\%} & \textbf{100.00\%} & 10.81\% & 19.51\% \\
    \midrule
    CrowdCanary & \cellcolor{yellow!25}\textbf{DomainDynamics} & \cellcolor{yellow!25}0.41\% & \cellcolor{yellow!25}99.71\% & \cellcolor{yellow!25}\textbf{94.02\%} & \cellcolor{yellow!25}\textbf{96.78\%} \\
    & Baseline & \textbf{0.00\%} & \textbf{100.00\%} & 5.95\% & 11.24\% \\
    \midrule
    ALL & \cellcolor{yellow!25}\textbf{DomainDynamics} & \cellcolor{yellow!25}0.41\% & \cellcolor{yellow!25}99.87\% & \cellcolor{yellow!25}\textbf{82.58\%} & \cellcolor{yellow!25}\textbf{90.40\%} \\
    & Baseline & \textbf{0.00\%} & \textbf{100.00\%} & 8.14\% & 15.06\% \\
    \bottomrule
    \end{tabular}
    }
\end{table*}

\subsection{Comparison with Baseline Systems}
\label{sec:baseline}
In comparison to DomainDynamics, this study evaluates the performance against previous research using the same dataset and metrics.

\noindent\textbf{DomainDynamics.}
DomainDynamics employs the optimal parameters identified in Section~\ref{sec:detection-performance-parameters} and reports the performance when evaluating Malware, Phishing, and CrowdCanary in the Predicting Dataset individually as well as their combined performance in Table~\ref{tab:performance-baselines}.
The results for DomainDynamics in Table~\ref{tab:performance-baselines} for ALL (Malware+Phishing+CrowdCanary) correspond to those when the ML Algorithm in Table~\ref{tab:performance} is set to XGBoost.
When comparing individual results for Malware, Phishing, and CrowdCanary in Table~\ref{tab:performance-baselines}, it is evident that the detection performance is superior in the order of Malware, Phishing, and CrowdCanary.
This indicates that the difficulty of assessing the risk of domain names varies depending on the type of attack.

\noindent\textbf{Baseline (DomainProfiler).}
First, we consider the comparison with previous research.
It is important to note, as described in Section~\ref{sec:background}, that DomainDynamics is the first to provide outputs that consider changes in risk, which was not possible in conventional research, hence a direct comparison with prior work is inherently limited.
Specifically, conventional studies have reported accuracy evaluation results by labeling domain names collected over a certain period as ``Malicious'' using blocklists, etc., even if they were no longer used in attacks after a certain point.
If predicted as malicious, they are considered true positives.
However, this study does not use such metrics for evaluation; instead, we assess and compare with our metrics (Section~\ref{sec:metrics}) centered on whether detection occurs prior to an attack.
Moreover, the previous studies mentioned in Table~\ref{table:list-related} are neither open source nor provide publicly available labeled data for constructing ML models, making them effectively inaccessible to third parties.
Therefore, we re-implemented DomainProfiler based on the description in the original paper~\cite{DBLP:conf/dsn/ChibaYASYMG16}, which is the most representative of the previous research with such conventional labeling, and applied it to our dataset to evaluate the performance of DomainProfiler.
We obtained the historical output of DomainProfiler for the corresponding domain names.
We then calculate the metrics based on these results.
The corresponding results can be found in the row for Baseline (DomainProfiler) in Table~\ref{tab:performance-baselines}.
As a result, DomainProfiler achieved an FPR of 0\% and a precision of 100\% on our dataset.
However, for ALL (Malware+Phishing+CrowdCanary), the Recall/TPR was only 8.14\%, and the F1 score was limited to 15.06\%.
It is evident that DomainProfiler was significantly less capable of detecting attacks before they occurred, compared to DomainDynamics with the current dataset.

\subsection{Real-world Deployment}
\label{sec:real-world}
We implemented DomainDynamics in the operational environment of a real-world security service provider, applying it to assess the risk of a vast number of domain names observed in a large-scale network and those newly registered in domain registries.

\noindent\textbf{Data Sources}
We acquired domain names from two types of data sources for risk assessment using DomainDynamics.
The first source was domain names observed through passive DNS traffic, collected from 66 DNS servers installed across 18 countries on global Tier 1 networks.
We obtained 55.57 million domain names that were newly resolved during the 28 days from November 27, 2022, to December 24, 2022, and assessed their risk in real-time with DomainDynamics.
The second source was newly registered domain names observed in domain registries, collected using Zonefiles.
Over the same 28-day period, we daily acquired newly registered domain names, totaling 3.86 million, and assessed their risk with DomainDynamics.

\noindent\textbf{Verification Method}
We verified domain names identified as risky by DomainDynamics using three methods.
First, we used the VirusTotal API to check daily for up to one month after detection by DomainDynamics whether those domain names were identified by any of the 88 detection engines.
Second, we verified whether the domain names were C2 domains generated on the day of detection by known DGAs using a commercial Threat Intelligence service utilized by the security service provider, checking daily for up to one month after detection.
Third, we conducted daily web crawls for up to one month after detection by DomainDynamics.
Whenever web access was achieved, we took screenshots to manually check whether the sites were phishing sites.
However, there are constraints: the first and second methods are challenging to conduct daily on all items due to the high licensing costs of the APIs used, and the third method is difficult to implement on a large scale and daily in an ethically considerate and scalable way.
Therefore, we randomly sampled 100,000 domain names detected by DomainDynamics and verified them daily for up to one month following detection.

\noindent\textbf{Verification Results}
First, as a result of verification using the VirusTotal API, out of the 100,000 domain names assessed for risk by DomainDynamics, 6,937 were detected by at least one of VirusTotal's engines after detection.
Although the number is not large due to the generally low detection rate of VirusTotal, it was confirmed that there indeed are domain names among those detected by DomainDynamics that are used in malicious activities and judged malicious.
Next, as a result of verification using Threat Intelligence, out of the 100,000 domain names assessed for risk by DomainDynamics, 16,674 were confirmed to be C2 domain names generated by DGAs and used after detection.
This result indicates that DomainDynamics can preemptively detect C2 domain names generated by DGAs.
Lastly, as a result of verification through web crawling, out of the 100,000 domain names assessed for risk by DomainDynamics, 1,552 were confirmed to be operating as phishing sites through screenshots taken after detection.
Despite the low frequency of web crawls and only accessing domains directly without specifying URL paths, DomainDynamics was shown to preemptively detect a certain number of phishing sites.

\section{Discussion}
\label{sec:discussion}

\subsection{Ethical Consideration}
\label{sec:ethical}
Our study raises no ethical concerns.
We did not use any user-related data.
In the deployment described in Section~\ref{sec:real-world}, we utilized Passive DNS data, which resides between cache and authority, containing information solely on the resolved domain names without revealing any data about the users who initiated the name resolutions.
Furthermore, our approach to web crawling in Section~\ref{sec:real-world} was carefully managed to avoid ethical concerns, including implementing a delay of at least three seconds between accesses to the same server.
Consequently, our organization did not view our research as involving human subjects, thus we did not seek IRB approval, in compliance with the ethical considerations outlined in our study.

\subsection{Limitation}
\label{sec:limitation}
We acknowledge several limitations of this research, which are outlined below.

\noindent\textbf{Potential for Evasion Detection.}
By disclosing the details of DomainDynamics, there is a risk that attackers might devise strategies to avoid leaving traces in historical data, thereby evading detection, or to alter the features that contribute to risk prediction.
Nevertheless, in the current Internet environment, it is practically impossible to prepare a domain name for user access without leaving any trace in the historical data of WHOIS records, SOA records, and TLS certificates used by DomainDynamics, without being utilized as features by our system.

\noindent\textbf{Difficulty in Inferring Attack Objectives.}
While DomainDynamics excels at assessing the risk level of a domain name at specific points in its lifecycle, it does not categorize the nature of the threat (e.g., malware, phishing).
This limitation stems from its methodological approach, which focuses on temporal risk assessment rather than detailed threat classification.
In practical terms, irrespective of the threat being malware or phishing, the critical factor is the ability to preemptively identify and neutralize potential harm by either blocking access to high-risk domain names or preemptively taking action against them.
Thus, the granular identification of attack types, while informative, is secondary to the primary goal of mitigating risk through lifecycle-based domain analysis.

\section{Conclusion}
\label{sec:conclusion}
In conclusion, this research presents DomainDynamics, an innovative system designed to enhance the detection of malicious domain names by leveraging a lifecycle-aware approach.
By constructing timelines for domain names and applying ML to analyze characteristics at different lifecycle stages, DomainDynamics effectively addresses the prevalent issues of false positives and false negatives in traditional detection methods.
The evaluation of DomainDynamics, conducted with a substantial dataset of over 85,000 known malicious domain names, has demonstrated its efficacy.
The system achieved an impressive detection rate of 82.58\% within a seven-day forecast period, while maintaining a low false positive rate of 0.41\%.
This performance significantly exceeds that of prior studies and commercial services, marking a notable advancement in cybersecurity.
Ultimately, DomainDynamics represents a significant step forward in proactive cybersecurity, offering a robust tool to predict and mitigate cyber threats more effectively and enhancing the overall resilience of digital infrastructures.

\bibliographystyle{IEEEtran}
\bibliography{main}

\end{document}